\documentclass[aps,prb,twocolumn,showpacs,10pt]{revtex4}
\usepackage[utf8]{inputenc}
\usepackage{graphicx,amsmath}
\usepackage{amssymb}
\usepackage{amsthm}

\begin{document}
\title{Equivalent topological invariants for one-dimensional Majorana wires in symmetry class D}
\author{Jan Carl Budich, Eddy Ardonne}

\affiliation{Department of Physics, Stockholm University, SE-106 91 Stockholm, Sweden}
\date{\today}
\begin{abstract}
Topological superconductors in one spatial dimension exhibiting a single Majorana bound state at each end are distinguished from trivial gapped systems by a $\mathbb Z_2$~topological invariant. Originally, this invariant was calculated by Kitaev in terms of the Pfaffian of the Majorana representation of the Hamiltonian: The sign of this Pfaffian divides the set of all gapped quadratic forms of Majorana fermions into two inequivalent classes. In the more familiar Bogoliubov de Gennes mean field description of superconductivity, an emergent particle hole symmetry gives rise to a quantized Zak-Berry phase the value of which is also a topological invariant. In this work, we explicitly show the equivalence of these two formulations by relating both of them to the phase winding of the transformation matrix that brings the Majorana representation matrix of the Hamiltonian into its Jordan normal form.  
\end{abstract}
\maketitle

\section{Introduction}
The topological superconductor in one spatial dimension (1DTSC) has been discovered and classified in a pioneering 2001 paper by Kitaev \cite{Kitaev2001}. This state of matter features a single isolated Majorana bound state (MBS) at each of its ends. The 1DTSC is also intriguing from a conceptual point of view since it is the only known topological phase in one dimension that is not symmetry protected. Not symmetry protected means that a non-trivial 1DTSC cannot be connected to a trivial state without going through a gap-closing phase transition. The original proposal considers an effective spinless p-wave superconductor (SC). More recently, the 1DTSC phase has also been identified in nanowires that are proximity coupled to an s-wave SC \cite{SauTSC,OppenTSC}. In these systems, the combination of Rashba spin orbit coupling and a Zeeman splitting is employed to dispose of the spin degree of freedom in the effective low energy theory. The first experimental signatures of MBS have been reported by several experimental groups \cite{LeoMaj, LarssonXu, HeiblumMaj}.\\

In recent years, there has been enormous interest in topological states of matter (TSM) that can be understood at the level of quadratic model Hamiltonians \cite{HasanKane,XLReview2010,TSMReview}. A complete classification of these TSM, one example of which is the 1DTSC in symmetry class D \cite{AltlandZirnbauer}, has been achieved by different means in Refs. \onlinecite{Schnyder2008,KitaevPeriodic,RyuLudwig}. Here, we would like to focus on fully gapped (proximity induced) SCs without any additional symmetries at mean field level. In the language of Ref. \onlinecite{KitaevPeriodic}, such systems are characterized by a quadratic form of Majorana operators without any physical symmetries. Note that due to the Majorana algebra, the representation matrix is automatically antisymmetric. In this framework, the 1DTSC has been distinguished from a trivial fully gapped system by the sign of the Pfaffian of this representation matrix \cite{Kitaev2001}. On the other hand, within the approach of Refs. \onlinecite{Schnyder2008,RyuLudwig}, the Bogoliubov de Gennes (BdG) mean field Hamiltonian of a SC is treated on the same footing as the Bloch bands of a non-interacting insulator. The Nambu spinor structure representing two copies of the actual excitation spectrum, a particle and a hole copy, is then reflected in the formal emergence of a particle hole symmetry (PHS) $\mathcal C$~with $\mathcal C^2=+1$. This antiunitary PHS implies a quantization of the Zak-Berry phase \cite{Berry,ZakPol} associated with a 1D band structure to integer multiples of $\pi$ \cite{HatsugaiQuantizedBerry}. Hence, there are only two distinct values, $0 ~\text{(mod 2$\pi$)}$~and $\pi~\text{(mod 2$\pi$)}$,~for the Zak-Berry phase which defines a $\mathbb Z_2$~invariant. We refer to Ref. \onlinecite{BernevigHughes} for a recent overview on the theory of charge polarization in 1D systems.\\

In this work, we would like to explicitly demonstrate the equivalence between these two approaches to the topological $\mathbb Z_2$-invariant characterizing the 1DTSC. To this end, we proceed in two steps. First, we express Kitaev's Pfaffian invariant as the phase winding between the two real points of the Fourier transform of the orthogonal transformation that brings the antisymmetric Majorana representation matrix of the mean field Hamiltonian into its Jordan normal form. Second, we start from the Berry connection associated with the BdG band structure and show that the quantized Berry phase can be expressed as the same phase winding. Interestingly, this equivalence implies that Kitaev's Pfaffian invariant can also be used as a convenient means to calculate the quantized Berry phase for a normal insulating 1D system with a physical PHS, e.g., an insulator similar to the model introduced by Su, Schrieffer, and Heeger (SSH) \cite{SSH,SSHReview}. For a chiral symmetry protected 1DTSC in symmetry class BDI, a complementary analysis has been presented in Ref. \onlinecite{SauChiral} (see also Ref.~\onlinecite{SatoChiral}). There, the parity of the winding number characterizing a chiral 1D system \cite{Ryu2002,RyuLudwig} is shown to be equal to Kitaev's Pfaffian invariant. In Ref. \onlinecite{z2maj}, the $\mathbb Z_2$~invariant characterizing the 1DTSC has been calculated in terms of its single particle Green's function using a dimensional extension to the 2DTSC which extends the domain of the invariant beyond the set of mean field Hamiltonians.\\

The remainder of this article is organized as follows: In Section \ref{sec:Kitaev}, we express Kitaev's Pfaffian invariant as the phase winding of a determinant between the real $k$-points $0$~and $\pi$. In Section \ref{sec:ZakBerry}, the quantized Berry phase of a 1D BdG band structure is shown to be given by the same phase winding, using the emergent PHS. Section \ref{sec:SSH} is dedicated to a discussion of non-superconducting systems with a physical PHS where the quantized Berry phase can also be expressed as the sign of the product of two Pfaffians evaluated at the real $k$-points. In Section \ref{sec:conclusion}, a concluding discussion is presented.

\section{Pfaffian invariant without Pfaffians}
\label{sec:Kitaev}
The Hamiltonian of a SC at mean field level on a lattice is a quadratic form in the field operators $\psi_j$, where $j$~labels the real space position on the lattice and $\psi_j=(\psi_{j,1},\ldots,\psi_{j,n})^T$~is a spinor comprising $n$~internal degrees of freedom like spin, orbital-label, etc. Without loss of generality we will always consider a lattice with unit lattice constant here. Since the mean field SC is not particle number conserving, the Hamiltonian is not of the form $\psi_i^\dag H_{ij}\psi_j$~but contains terms of the form $\psi_i \Delta_{ij} \psi_j + \text{h.c.}$~with some pairing matrix $\Delta$. A convenient formalism to study the properties of such generalized quadratic Hamiltonians is to go to a basis of (spinors of) Majorana fermions $\gamma_{j,x}=\psi_j+\psi_j^\dag,~\gamma_{j,y}=-i(\psi_j-\psi_j^\dag)$. Here, the labels $x,y$~on the Majorana operators allude to the real ($x$) and imaginary ($y$) part of the complex fermion $\psi_j$. The real Majorana operators satisfy the algebra
\begin{align}
&\gamma_{j,a}^\dag=\gamma_{j,a},\quad a=x,y\nonumber\\
&\left\{\gamma_{i,a},\gamma_{j,b}\right\}=2\delta_{i,j}\delta_{a,b}.
\end{align}
For notational brevity, we will suppress the $a=x,y$~index, by defining the spinors $\gamma_j=\left(\gamma_{j,x},\gamma_{j,y}\right)^T$. It then follows that any mean field Hamiltonian (SC or not) can be expressed in the form \cite{KitaevPeriodic}
\begin{align}
H=\frac{i}{2}\sum_{i,j}\gamma_i^T A_{ij}\gamma_j ,
\end{align}
which is called the Majorana representation. The representation matrix $A$~is real and antisymmetric. Hence, it has purely imaginary eigenvalues that occur in complex conjugate pairs $\pm i\epsilon_\lambda,~\epsilon_\lambda>0$. We note that we consider a gapped system with periodic boundary conditions, in which case there are no zero modes. By means of a real orthogonal transformation $W$~such a matrix can be brought into the block-diagonal Jordan form
\begin{align}
A_J = WAW^T = \text{diag}_\lambda\begin{pmatrix}
                          0 & \epsilon_\lambda\\{-\epsilon_\lambda}&0
                         \end{pmatrix}.
\label{eqn:jordan}                         
\end{align}
The transformation matrix $W$~is here defined up to a global sign which does not change the determinant of $W$~due to the even dimension of the real vector space on which $W$~acts. This is true independent of the system size since there are always two Majorana fermions per complex degree of freedom.   
\\

Kitaev\cite{Kitaev2001} introduces the notion of the `Majorana number' $\mathcal{M}(H)=\pm 1$
associated with the Hamiltonian $H$. For Hamiltonians that exhibit a MBS in the case of open boundary conditions, $\mathcal{M}(H)$ takes the value $-1$, which corresponds to a non-trivial 1DTSC.

Kitaev then relates $\mathcal{M}(H)$
to the fermionic parity of the ground state of a closed chain of length $L$, denoted by
$P\bigl(H(L)\bigr)$, in the following way
\begin{equation}
\label{eq:mn}
P\bigl(H (L_1+L_2) \bigr) = \mathcal{M}(H) P\bigl(H (L_1) \bigr) P\bigl(H (L_2) \bigr) \ .
\end{equation}
For general non-interacting systems, the fermionic parity of the ground state of a Hamiltonian
$H$ can be shown to take the form
\begin{equation}
\label{eq:ph}
P (H) = \text{sgn}\left\{\text{Pf}\left(A\right)\right\} \ ,
\end{equation}
where $\text{Pf}\left(A\right)$ denotes the Pfaffian of the anti-symmetric matrix
$A$, given in terms of the totally anti-symmetric tensor $\varepsilon_{i_1,i_2,\ldots,i_{2n}}$ as
\begin{equation}
\text{Pf}\left(A\right) =
\frac{1}{2^n n!} \varepsilon_{i_1,i_2,\ldots,i_{2n}} A_{i_1,i_2} A_{i_3,i_4}\cdots A_{i_{2n-1},i_{2n}} \ .
\end{equation}
For this reason, $\mathcal{M}$ is also referred to as the `Pfaffian' $\mathbb Z_2$~invariant.

We would now like to bring the Pfaffian $\mathbb Z_2$~invariant $\mathcal M$ \cite{Kitaev2001} characterizing the 1DTSC into a form which will allow us to make its equivalence to a quantized Zak-Berry phase manifest. Generally speaking, in differential topology, one is concerned with topological invariants associated with smooth manifolds. For a periodic system in the thermodynamic limit, the $k$-space is a smooth manifold on which the invariants of all TSM are defined \cite{RyuLudwig,TSMReview}. For the invariant defining the 1DTSC, the so called real $k$-points $k=0,\pi$ where $k=-k$~will be of crucial importance. When doing a numerical calculation, one is sometimes forced to consider finite system sizes. In this case, the real $k$-point $\pi$~only exists if the number of lattice sites is even. In agreement with Ref. \onlinecite{Kitaev2001}, we hence expect that the analytical form of the invariants in $k$-space can only be extended to finite systems with an even number of lattice sites
which we will assume in the following.

It is then an immediate consequence of Eqs. \eqref{eq:mn} and \eqref{eq:ph} that the topological invariant $\mathcal M$~is simply given by
\begin{align}
\mathcal M=\text{sgn}\left\{\text{Pf}\left(A\right)\right\} .
\label{eqn:defM}
\end{align}

Defining the Fourier transform of the Majorana representation matrix of a translation invariant system as $\tilde A(k)$, Eq. (\ref{eqn:defM}) can be expressed as
\begin{align}
\mathcal M= \text{sgn}\left\{\text{Pf}\left(\tilde A(0)\right)\text{Pf}\left(\tilde A(\pi)\right)\right\},
\label{eqn:KitaevInvariant}
\end{align}
which is the probably best known form of the invariant involving, as already mentioned, the real $k$-points $k=0$~and $k=\pi$.\\

The Pfaffian of the Jordan form $A_J$~(see Eq. (\ref{eqn:jordan})) is easy to evaluate:
\begin{align}
\text{Pf}\left(A_J\right)=\prod_\lambda \epsilon_\lambda~>0.
\end{align}
Using the elementary algebraic relation $\text{Pf}\left(WAW^T\right)=\text{Pf}\left(A\right) ~\text{det}\left(W\right)$~along with Eq. (\ref{eqn:defM}), we immediately get
\begin{align}
\mathcal M = \text{det}\left(W\right) = \pm 1.
\end{align}
For a translation invariant system, the Fourier transform $\tilde W(k)$~of $W$~is block diagonal and we get
\begin{align}
&\mathcal M= \prod_k \text{det}\left(\tilde W(k)\right)=\prod_{k=-k}\text{det}\left(\tilde W(k)\right)=\nonumber\\
&\text{det}\left(\tilde W(0)\right)\text{det}\left(\tilde W(\pi)\right).
\end{align}
The second equality sign here uses the reality of $W$~which implies $\tilde W(k)^*=\tilde W(-k)$. Since $\tilde W(k)$~is unitary for all $k$, its determinant can be written as $\text{det}\left(\tilde W(k)\right)=\text{e}^{i\varphi_k}$. The reality constraint then yields $\varphi_k=-\varphi_{-k} ~(\text{mod 2}\pi)$~implying that $\varphi_k$~is quantized to integer multiples of $\pi$~at the real $k$-points. The invariant can hence be expressed as
\begin{align}
\mathcal M=(-1)^{\frac{\varphi_0-\varphi_\pi}{\pi}}.
\label{eq:mdeltaphi}
\end{align}
The determinant of $\tilde W(k)$~is a continuous function of $k$~so the phase change $\Delta \varphi=\varphi_0-\varphi_\pi$~can be written as the following `winding' integral over half of the Brillouin zone
\begin{equation}
\Delta \varphi = i\int_0^\pi\left[\partial_k \left(\log\det\left(\tilde W(k)\right)\right)\right] \text{d}k.\\
\label{eqn:phasewind}
\end{equation}
In the next Section, we will derive the same expression for the quantized Berry phase associated with the BdG band structure of a mean field SC.

\section{Relation between quantized Berry phase and Pfaffian invariant}
\label{sec:ZakBerry}
Above, we worked with the Majorana representation of SC mean field Hamiltonians. We now tie the topological invariant $\mathcal{M}$ obtained in the Majorana representation to the more widely used BdG picture. In the BdG picture, the Hamiltonians are represented in the Nambu spinor basis $\Psi_j=(\psi_j,\psi_j^\dag)^T$ as follows
\begin{align}
H=\sum_{i,j}\Psi_i^\dag \left(H_{\text{BdG}}\right)_{ij}\Psi_j.
\end{align}
In contrast to the Majorana operators $\gamma_{j,x},~\gamma_{j,y}$~that span the same Hilbert space as $\psi_j,\psi_j^\dag$, the Nambu basis is not real and neither is the representation matrix $H_{\text{BdG}}$. The interdependence of $\psi_j$~and $\psi_j^\dag$~is then reflected in an `emergent' PHS $\mathcal C=\tau_x K$, i.e., $\left\{H_{\text{BdG}},\mathcal C\right\}=0$,
with $\mathcal{C}^2 = 1$. As a consequence, the BdG band structure in the absence of further symmetries is in the Altland-Zirnbauer \cite{AltlandZirnbauer} symmetry class D.
Here, $K$~denotes the complex conjugation and $\tau_x$~is a Pauli matrix acting in the Nambu space. The approach of Refs. \onlinecite{Schnyder2008,RyuLudwig} to the topological classification is to treat the Fourier transform $\tilde H_{\text{BdG}}(k)$~of the BdG Hamiltonian on the same footing as the Bloch Hamiltonian of an ordinary insulator with a physical PHS. The corresponding $\mathbb Z_2$~topological invariant is then the 1D Chern-Simons invariant, i.e., a Zak-Berry phase that is quantized to integer multiples of $\pi$~due to the antiunitary PHS. This symmetry also implies that the bands below and above the energy gap are not independent, but are conjugated by PHS. We employ this dependence defining
\begin{align}
\mathcal C\lvert u_\alpha^o(-k)\rangle=\text{e}^{-i\chi_\alpha(k)}\lvert u_\alpha^e(k)\rangle,
\label{eqn:phsBloch}
\end{align}
where $\alpha=1,\ldots, n$~labels the independent Bloch bands and $\lvert u_\alpha^o\rangle,~(\lvert u_\alpha^e\rangle)$~denotes the Bloch states associated with the occupied (empty) bands. Using Eq. (\ref{eqn:phsBloch}), one can show that the Abelian Berry connection $\mathcal A^o(k)=-i\sum_\alpha \langle u_\alpha^o(k)\lvert \partial_k\rvert u_\alpha^o(k)\rangle$~associated with the occupied bands is related to the similarly defined $\mathcal A^e(k)$~by (in \cite{FUPump}, an analogous relation was used in the context of time-reversal invariant systems) 
\begin{align}
\mathcal A^o(-k)=\mathcal A^e(k)-\sum_\alpha \partial_k\chi_\alpha(k),
\label{eqn:constraintBerry}
\end{align}
i.e., the Berry connection of the occupied bands at $k$~is the Berry connection of the empty bands at $-k$~up to a gauge transformation. With this constraint the Zak-Berry phase $\Phi_{\text{ZB}}$~can be simplified to an integral over half of the Brillouin zone as
\begin{align}
\Phi_{\text{ZB}}=\int_{-\pi}^{\pi}\mathcal A^o(k)\text{d}k=\int_0^\pi\left[\mathcal  A(k)-\sum_\alpha\partial_k\chi_\alpha(k)\right]\text{d}k,
\label{eqn:defBerry}
\end{align}
where the first equality sign defines the Zak-Berry phase of the gapped system and Eq. (\ref{eqn:constraintBerry}) along with the definition $\mathcal A(k)=\mathcal A^o(k)+\mathcal A^e(k)$~enters the second equality sign.\\

To make further progress, we take a look at the relation between the Majorana representation and the BdG picture. The Majorana spinor $\gamma_j=(\gamma_{j,x},\gamma_{j,y})^T$~and the Nambu spinor $\Psi_j=(\psi_j,\psi_j^\dag)^T$~are related by the unitary transformation $U$~like
\begin{align}
\gamma_j=\sqrt{2}U \Psi_j,\quad U=\frac{1}{\sqrt{2}}\begin{pmatrix}
                                             1&1\\{-i}&i
                                            \end{pmatrix}.
\end{align}
Hence, the corresponding representation matrices are related as $i A=U H_{\text{BdG}}U^\dag$. Since $U$~commutes with the Fourier transform, there is a similar relation in $k$-space, explicitly $i \tilde A(k)=U\tilde H_{\text{BdG}}(k)U^\dag$. Going to the Jordan form $A_J$~of the Majorana representation matrix, we note that $U$~brings us to the diagonal form of the BdG-Hamiltonian, i.e.,
\begin{align}
i U^\dag A_J U= \text{diag}_\lambda\begin{pmatrix}
                           \epsilon_\lambda&0\\0&{-\epsilon_\lambda}
                         \end{pmatrix}=\tau_z \text{diag}_\lambda(\epsilon_\lambda).
\end{align}
Defining $\hat W(k)=U^\dag\tilde W(k) U$~we hence obtain
\begin{align}
\hat W(k)\tilde H_{\text{BdG}}(k) \hat W^\dag(k)=\tau_z ~\text{diag}\left(\epsilon_1(k),\ldots,\epsilon_n(k)\right) .
\end{align}
with the energy eigenvalues $\epsilon_\alpha(k)>0,~\alpha=1,\ldots,n$~of the empty Bloch bands. This allows us to define a global gauge for the Bloch states $\lvert u_\alpha^\sigma(k)\rangle,~\sigma=o,e$~as follows:
\begin{align}
\lvert u_\alpha^\sigma(k)\rangle=
\sum_\beta \hat W^*_{\alpha,\beta}(k)\lvert \beta,\sigma\rangle,
\end{align}
where $\lvert\alpha,\sigma\rangle=\frac{1}{\sqrt{L}}\sum_j \Psi_{j,\alpha,\sigma}^\dag\lvert \text{vac}\rangle$~is the lattice periodic part of the canonical basis states $\frac{1}{\sqrt{L}}\sum_j\text{e}^{ikj}\Psi_{j,\alpha,\sigma}^\dag\lvert \text{vac}\rangle$ associated with the Nambu-spinor basis in $k$-space (where $\sigma$ denotes the Nambu index).

In this gauge, the phase factors $\chi_\alpha$~ appearing in Eq.~\eqref{eqn:phsBloch} vanish. This is  because of PHS, which in the Majorana representation implies $K \tilde{W}(k) K = \tilde{W}(k)^* = \tilde{W}(-k)$, or, in the
current BdG basis,
\begin{align}
\mathcal C \hat W(k) \mathcal C^{-1}=\hat W(-k).
\end{align}
From this equation, it directly follows that the phase factors $\chi_\alpha (k)$ in
Eq.~\eqref{eqn:phsBloch} vanish. Using Eq. (\ref{eqn:defBerry}), we can hence write
\begin{align}
&\Phi_{\text{ZB}}=i\int_0^\pi \text{Tr}\left[\hat W^\dag(k)\partial_k \hat W(k)\right]\text{d}k=\nonumber\\
&i\int_0^\pi\left[\partial k \left(\log\det\left(\tilde W(k)\right)\right)\right] \text{d}k= \Delta\varphi,
\end{align}
where Eq. (\ref{eqn:phasewind}) has been used for the last equality sign. This makes the equivalence of the expressions (\ref{eqn:KitaevInvariant}) and (\ref{eqn:defBerry}) manifest which was the main purpose of the present analysis.
We note that due to the $\pi$-quantization of the Berry phase
$\Delta\varphi=-\Delta\varphi~ (\text{mod 2}\pi)$.
This is reflected in the expression Eq.~\eqref{eq:mdeltaphi} for $\mathcal{M}$, which
does not depend on the overall sign of $\Delta\varphi$.

\section{Pfaffian invariant for non-superconducting systems with PHS}
\label{sec:SSH}
In the 1DTSC phase, PHS is not a physical symmetry but `emerges' from the BdG description of superconductivity. However, there are also symmetry protected topological states in particle number conserving 1D systems that have a physical PHS $\mathcal C$~with $\mathcal C^2=1$. The most prominent example of this category is the SSH model \cite{SSH,SSHReview}. In this case, the defining $\mathbb Z_2$~invariant is not associated with the presence of a single MBS but with a localized fermionic state with a fractional charge of $\frac{e}{2}$ \cite{Jackiw2012}. However, a Pfaffian invariant can also be defined in this case as we will discuss now. Generally, the operation of PHS can be expressed as $\mathcal C=U_{\mathcal C}K$~with the unitary part $U_{\mathcal C}$. Under a unitary transformation $U$~this unitary part transforms like $U_{\mathcal C}\rightarrow UU_{\mathcal C}U^T$~due to the complex conjugation involved in $\mathcal C$. In particular, because $\mathcal{C}^2=1$ implies that $U_\mathcal{C} = U_\mathcal{C}^T$, one can always find a unitary $U$~that satisfies $UU_{\mathcal C}U^T=1$.
After this transformation, $\mathcal C=K$, implying that the transformed Hamiltonian is of the form $UHU^\dag=iA$, where $A$~is a real antisymmetric matrix. Although the basis vectors in this representation are not Majorana fermions, it is formally completely analogous to the Majorana representation of SC mean field Hamiltonians. Hence, a Pfaffian invariant can be defined in terms of the antisymmetric matrix $A$~identical to Eq. (\ref{eqn:defM}) and Eq. (\ref{eqn:KitaevInvariant}), respectively. The proof that this invariant is equal to the quantized Zak-Berry phase, which is well known to topologically classify PHS protected topological states in 1D, is analogous to the one for the superconducting case presented above. We note that evaluating the Pfaffian invariant can be much more convenient as it does not involve an integration over the Brillouin zone but only contains information about the Bloch Hamiltonian at the two real $k$-points.

\section{Concluding discussion}
\label{sec:conclusion}
We have made the equivalence manifest between the quantized Zak-Berry phase and the Pfaffian invariant characterizing a 1DTSC and a PHS conserving 1D insulator in symmetry class D, respectively. This has been achieved by expressing both formulations as the phase winding of the determinant of a unitary matrix in half of the Brillouin zone. The other half of the Brillouin zone is redundant due to the antiunitary constraint of PHS. The equivalence between the two approaches to the topological invariant is not limited to superconducting systems but also holds for symmetry protected topological states in 1D like the SSH model.  Our construction is related to a similar analysis\cite{FUPump} (see also Ref.~\onlinecite{RoyZ2}) of two-dimensional systems in the symplectic symmetry class AII , where the relevant $\mathbb Z_2$~invariant could be connected to a so called time reversal polarization. In 1D, the quantized Zak-Berry phase is well known to correspond to a polarization of the underlying lattice and so does the equivalent Pfaffian invariant. However, for the BdG band structure, this polarization is in general not a charge polarization that has immediate observable consequences.\\   

{\em Acknowledgements}.
We would like to thank Hans Hansson for numerous enlightening discussions. 
This research was sponsored, in part, by the swedish research council.

%\bibliographystyle{apsrev}
%\bibliography{kb}

\end{document}